\documentclass[superscriptaddress,twocolumn,showpacs,floatfix,prl]{revtex4}

\usepackage{amsmath}
\usepackage{graphicx}

\begin{document}

\title{Universal dynamic scaling in three-dimensional Ising spin glasses}

\author{Cheng-Wei Liu}
\affiliation{Department of Physics, Boston University, 590 Commonwealth
Avenue, Boston, MA 02215, USA
}

\author{Anatoli Polkovnikov}
\affiliation{Department of Physics, Boston University, 590 Commonwealth Avenue,
Boston, MA 02215, USA
}

\author{Anders W. Sandvik}
\affiliation{Department of Physics, Boston University, 590 Commonwealth
Avenue, Boston, MA 02215, USA
}

\author{A.~P.~Young}
\affiliation{Department of Physics, University of California, Santa Cruz,
California 95064, USA}


\begin{abstract}

We use a non-equilibrium simulation method to study the spin glass transition in three-dimensional Ising spin glasses. The 
transition point is repeatedly approached at finite velocity $v$ (temperature change versus time) in Monte Carlo 
simulations starting at a high temperature. The normally problematic critical slowing-down is not hampering this kind of approach, 
since the system equilibrates quickly at the initial temperature and the slowing-down is merely reflected in the dynamic scaling of 
the non-equilibrium order parameter with $v$ and the system size. The equilibrium limit does not have to be reached. For the dynamic 
exponent we obtain $z = 5.85(9)$ for bimodal couplings distribution and $z=6.00(10)$ for the Gaussian case, thus supporting universal 
dynamic scaling (in contrast to recent claims of non-universal behavior).
\end{abstract}


\maketitle

Understanding spin glasses analytically has proved difficult, and there are only very few exact results beyond Parisi's 
solution~\cite{parisi:79, parisi:80, parisi:83} of the infinite-range Sherrington-Kirkpatrick model~\cite{sherrington:75}.
Furthermore, Monte Carlo (MC) simulations in three dimensions proved challenging because $d=3$ turns out to be close to the 
lower critical dimension $d_l$ below which fluctuations destroy the transition. For Ising spins, which we study here, 
$d_l\simeq 2.5$ \cite{boettcher:05}. Nonetheless, there has been substantial progress in recent years, aided by increased 
computer power, the technique of parallel tempering~\cite{hukushima:96} (exchange MC) to speed up equilibration and
reduce autocorrelations, and better methods of doing finite-size scaling. In particular, Hasenbusch et al.~\cite{hasenbusch:08} 
extracted not only the leading singular behavior at the transition but also the dominant correction to scaling. This gives 
confidence that the asymptotic critical region has been reached (which it had not in much of the earlier work, see, e.g., 
discussion in Ref.~\cite{katzgraber:06}), and hence that the critical exponents are accurate. Subsequently, massive 
simulations by Baity-Jesi et al.~\cite{baity-jesietal:13}, using a special-purpose, obtained even more accurate results 
which are consistent with the earlier work of Ref.~\cite{hasenbusch:08}.

In spite of these impressive developments, it is still useful to employ new techniques for studying spin 
glasses and other similar computationally hard problems. We here explore MC 
simulations out of equilibrium in combination with a powerful dynamic scaling theory \cite{liu:14} building on the Kibble-Zurek 
(KZ) mechanism \cite{kibble:76,zurek:85,janssen:89,zhong:05,polkovnikov:05,zurek:05,dziarmaga:05,dziarmaga:10,polkovnikov:11}. We will
demonstrate that this approach is particularly well suited for studies of spin glasses, because it circumvents the normally very problematic 
slowing-down (strong divergence in the MC autocorrelation time) at the transition, by turning it into a generic feature of the scaling 
methodology. Rapidly equilibrating a system at a high starting temperature, slow dynamics upon approaching the
transition is just reflected in the evolution of computed quantities with the velocity at which the temperature is changed. 
This process is described by scaling behaviors generalizing finite-size scaling. We here study three dimensional (3D) Ising spin glasses 
and address the issue of universal dynamics.

``Universality'' is a cornerstone of the theory of critical phenomena by which critical exponents and many other quantities 
do not depend on microscopic system details. Thus, the exponents for a spin glass should not depend on the distribution 
of interactions, so results obtained for, e.g., a bimodal distribution should be the same as those from a continuous distribution such 
as Gaussian. The work of Refs.~\cite{hasenbusch:08} and \cite{baity-jesietal:13} used a bimodal distribution of 
nearest-neighbor interactions, because considerable speedup in the MC code can be obtained in this case. An interesting question, 
raised by Campbell and collaborators~\cite{mari:99,mari:01,pleimling:05}, is whether universality may be violated in spin glasses.
While these works claim numerical evidence that the exponents \textit{do} depend on the distribution of interactions, other 
works, e.g., Ref.~\cite{katzgraber:06}, found universal behavior, though with some inconsistencies due to corrections to scaling 
not being incorporated.  Simulations of a quality comparable to that in Hasenbusch et al.~\cite{hasenbusch:08} 
and \cite{baity-jesietal:13}, which \textit{do} systematically incorporate the leading correction to scaling, have not yet been done 
for a continuous distribution of the interactions. Such simulations would show, beyond reasonable doubt, whether 
universality is satisfied in spin glasses.

The above discussion applies to static exponents. There is also considerable interest in the dynamics of spin glasses, since 
experimental spin glasses never equilibrate below the transition temperature $T_c$, and even as $T_c$ is approached from above relaxation 
times increase much more rapidly than in, say, ferromagnets. Several estimates of the dynamical exponent $z$ have been obtained 
and the results are summarized in Table~\ref{Tab:zvalues}. It is seen that there there are significant differences in the results from 
different works, and claims are again made~\cite{mari:01,pleimling:05} that the dynamical exponent, like the static ones, depend on the 
distribution of interactions. Using the non-equilibrium approach, which in recent applications to systems without disorder have proved
reliable in extracting the dynamic exponent \cite{liu:14}, we will show here that Ising spin glasses with bimodal and Gaussian 
distributions show the same universality (within small error bars).

\begin{table}
\begin{tabular*}{\columnwidth}{@{\extracolsep{\fill}} | l | c | c |}
\hline
\hline
Study & Model~ & Exponent $z$ \\
\hline
Pleimling and Campbell (Ref.~\cite{pleimling:05}) & $\pm J$ & $5.7(2)$\\
& G & $6.2(1)$ \\
Nakamura (Ref.~\cite{nakamura:06})$^\star$ & $\pm J$ & $5.1(1)$ \\
Katzgraber and Campbell (Ref.~\cite{katzgraber:05b})$^\star$ & G & $6.80(15)$\\
Rieger (Ref.~\cite{rieger:93})$^\star$ & $\pm J$ & $\simeq 6$\\
Ogielski (Ref.~\cite{ogielski:85}) & $\pm J$ & $6.0(8)$ \\
Belletti et al. (Ref.~\cite{bellettietal:09})$^\star$ & $\pm J$ & $6.86(16)$\\
This study & $\pm J$ & $5.85(9)$\\
           & G & $6.00(10)$\\
\hline
\hline
\end{tabular*}
\caption{
Estimates of the dynamical critical exponent $z$ for 3D Ising spin glasses with local updates (Metropolis dynamics) with 
a bimodal ($\pm J)$ or Gaussian coupling distribution (G). The papers indicated by an asterisk determine a 
non-equilibrium coherence length $\xi(t)$ below or at $T_c$. This increases with time $t$ like $t^{1/z_\text{eff}(T)}$ 
where an effective exponent $z_\text{eff}(T)$ is found empirically to vary as $T^{-1}$, and is also found to merge smoothly into the 
critical exponent $z$ at $T_c$, i.e.~$z_\text{eff}(T) = (T_c/T) z$. The value of $z$ quoted by the authors was obtained by using the best accepted 
value of $T_c$ at that time.  Reference \cite{rieger:93} plots values for a temperature dependent $x(T)$, related to $z_\text{eff}(T)$ by 
$x(T) = (d-2+\eta)/2z_\text{eff}(T)$. The value quoted in the table is obtained from the data point for $x(T)$ closest to the currently accepted 
$T_c$ and  $\eta$ values \cite{baity-jesietal:13}.}
\label{Tab:zvalues}
\end{table}

\textit{Spin glass models}.---We study Ising spin glasses that can be described by the Hamiltonian:
\begin{equation}
\mathcal{H} = \sum_{\langle i,j \rangle} J_{ij} \sigma_i \sigma_j,
\label{eq:Hamiltonian}
\end{equation}
where $\sigma_i \in \{-1,1\}$ and
$\langle i,j \rangle$ stands for the nearest neighbors on a simple cubic lattice. We consider (i) a bimodal distribution 
in which $J_{ij} = \pm 1$ with equal probability and (ii) $J_{ij}$ drawn from a Gaussian with mean $0$ and standard deviation $1$. 
The relevant quantity to characterize the spin glass transition is the Edward-Anderson order parameter:
\begin{equation}
q = \frac{1}{N} \sum_{i=1}^N \sigma_i^{(1)} \sigma_j^{(2)},
\label{eq:ea}
\end{equation}
where $(1)$ and $(2)$ stand for two independent simulations (``replicas'') of the same coupling realizations. We will study the mean squared order parameter 
$\langle q^2\rangle$. 

The spin glass transition temperature 
$T_c$ for the bimodal case has been determined to very high numerical accuracy \cite{hasenbusch:08b, baity-jesietal:13}; $T_c = 1.102(3)$. $T_c$ for the Gaussian 
case is not as well determined, although a reasonable estimate is also available \cite{marinari:98, katzgraber:05b}; $T_c = 0.94(2)$. The static exponents for the 
bimodal case are also well studied in \cite{hasenbusch:08, baity-jesietal:13}, which gave the correlation length exponent $\nu = 2.562(42)$ and correlation 
function exponent $\eta=-0.3900(36)$.

\textit{Dynamic simulation scheme}.---We perform MC simulation with the standard Metropolis algorithm on systems of linear size $L$ (number of 
spins $N=L^3$). Simulations start from an initial temperature $T_i = 2$, roughly twice $T_c$ where the system is easily equilibrated prior to each
dynamic (``quench'') simulation. We proceed with a linearly varying $T$ as a function of the simulation time $\tau$ (measured in units of a standard 
MC sweep consisting of $N$ spin flip attempts) until a final temperature $T_f = 0.5$ is reached. Thus, our quench velocity is defined as $v=1.5/\tau$ and
the temperature is lowered by $\Delta_T = v$ after each MC step. We choose the total quench time $\tau=150 \times 2^n$ with $n=0,1,2,\dots$. This
kind of process is also known as {\it simulated annealing} \cite{kirkpatrick:83}, but in that case one normally has in mind a very slow reduction of $T$ with 
the goal of finding an energy minimum or reaching equilibrium. We are interested in both slow and fast processes and carry out detailed studies of the behavior 
of averages over many quenches as a function of $v$ and $L$.

We use $64$-bit multi-spin coding for 64 replicas in a single run (using different random numbers for the acceptance probabilities 
for each replica, to avoid correlations), and when computing the order parameter (\ref{eq:ea}) we consider
overlaps between $32$ replica pairs. Since the fluctuations among different realizations of $J_{ij}$ will in general be much 
larger than the statistical errors within a given realization, we only perform one such 64-replica quench for each disorder realization. For 
small sizes and short quenches, we generated $\mathcal{O}(10^5)$ realizations and for larger sizes and longer quenches we have at least $\mathcal{O}(10^2)$ 
realizations. 

For simplicity of notation, we use $\langle \dots \rangle$ to denote all averages involved. After the simulations 
we use polynomial interpolation to obtain $\langle q^2 \rangle $ at any $T$ within $[T_i,T_f]$. We focus 
on the squared order parameter $\langle q^2 \rangle$ at or in the close vicinity of the known $T_c$. An alternative then would be to perform quenches 
to exactly $T_c$ (instead of continuing below $T_c$). However, we will also consider the propagation of errors from uncertainties of $T_c$, 
and in principle results below $T_c$ can also be used in further analysis of the spin glass state.

\textit{Dynamic scaling}.---According to well established equilibrium finite-size scaling theory \cite{barber:83}, the critical order parameter 
exactly at $T_c$ depends on the system size as
\begin{equation}
\langle q^2 \rangle_{eq} \sim L^{-2\beta/\nu} \sim L^{-(1+\eta)}.
\end{equation}
A recent study \cite{liu:14} building on the KZ mechanism \cite{kibble:76,zurek:85} and generalized dynamic finite-size 
scaling \cite{kibble:76,zurek:85,janssen:89,zhong:05,polkovnikov:05,zurek:05,dziarmaga:05,dziarmaga:10,polkovnikov:11} suggests that the order 
parameter at a continuous transition exhibits three different scaling regimes depending on $v$ and $L$ when quenching to $T_c$ (i.e., evaluating 
$\langle q^2\rangle$ at $T_c$ following a quench and averaging over quenches). We refer to Ref.~\cite{liu:14} for background and derivations 
and here only quote the final result:
\begin{equation}
  \langle q^2 \rangle = \left \{
\begin{array}{l r}
L^{-(1+\eta)} f_1 (v L^{z+1/\nu}),& v \lesssim v_{_{KZ}}(L), \\
L^{-d} v^{-x},  & v_{_{KZ}}(L)\ll v \ll 1, \\ 
L^{-d}  f_2 (1/v),   &  v \agt v_{_{KZ}}(L),
\end{array}
\right.
\label{eq:dual_scaling}
\end{equation}
where $d$ is the dimensionality and $v_{_{KZ}}(L) \sim L^{-(z+1/\nu)}$ is the characteristic KZ velocity that separates the adiabatic and non-adiabatic regimes. 
The first scaling  function $f_1$ governs the low-velocity regime and $f_2$ describes the high-velocity regime. There is a wide region $v_{_{KZ}}(L)\ll v \ll 1$
in which both functions reduce to a universal power-law behavior with the power $x$ related to the standard critical exponents:
\begin{equation}
x = (d-2\beta/\nu)(z+1/\nu)^{-1}.
\label{eq:power}
\end{equation}
The dynamic finite-size scaling forms (\ref{eq:dual_scaling}) have been thoroughly tested on standard Ising models \cite{liu:14} and yielded 
high-precision results for $z$ for different types of dynamics (local and cluster updates) and dimensionality. 

In the case of a glass, in particular, as anticipated in Ref.~\cite{liu:14} and demonstrated with results in the present work, a major additional advantage
of the quench approach combined with dynamic scaling is that uncertainties related to poor equilibration due to critical slowing down are avoided. In standard 
approaches one has to make sure that equilibrium indeed has been reached, and this can be very difficult to confirm in practice. In our approach, 
equilibration only has to be carried out at the high initial temperature $T_i$ (or, one could also start with some other initial condition). In the subsequent quench
process, equilibration, or lack thereof, is manifested as the scaling behaviors in Eq.~(\ref{eq:dual_scaling}), and the simulation results themselves
are never questionable. To study scaling one of course still has to reach low enough $v$ for the scaling function $f_2$ to cross over into the
asymptotic power-law form, but this occurs much before equilibrium is reached and much larger system sizes can be studied reliably than in
standard approaches.

\begin{figure}
\includegraphics[width=7cm, clip=true]{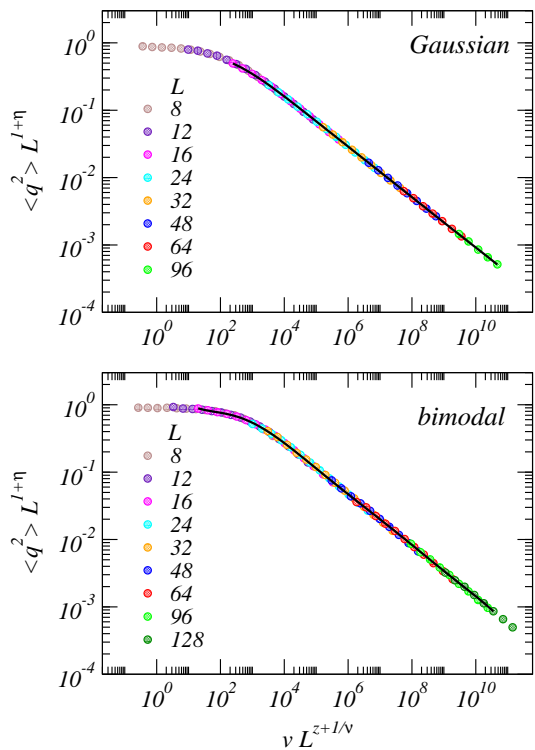}
\caption{(Color online) Scaling after quenching to $T_c$, using the form with $f_1$ in Eq.~(\ref{eq:dual_scaling}). The only free parameter 
optimized for scaling collapse is the dynamic exponent $z$, for which we obtain $z=5.85(9)$ and $6.00(10)$ for bimodal (top) and Gaussian (bottom) 
distributions, respectively.}
\label{fig1}
\end{figure}

\textit{Simulation results}.---We use many different velocities and system sizes $L=8, 12, \dots$ up to $L=128$ 
for the bimodal case and up to $L=96$ for the Gaussian case. With the static exponents $\nu$ and $\eta$ known (we use the values quoted above for both 
the bimodal and Gaussian cases), the dynamic exponent $z$ enters as the only unknown parameter in Eq.~(\ref{eq:dual_scaling}). One can treat $z$ as an 
adjustable parameter for optimal scaling collapse according to either the low-velocity (function $f_1$) or high-velocity ($f_2$) form in 
Eq.~(\ref{eq:dual_scaling}). We quantify the quality of the collapse using $\chi^2$ per degree of freedom relative to a function fitted to all 
the data, using a line in the linear regime on the log-log scale and matching it to a high-order polynomial describing the deviations from this 
form. After $z$ is determined this way, we introduce Gaussian noise to $T_c$, $\nu$, and $\eta$ with standard deviation equal to error bars quoted, repeating 
the scaling analysis with such altered data many times to obtain error estimates for $z$.

Results when using the first of Eq.~(\ref{eq:dual_scaling}), which delivering the scaling function $f_1$ when the data collapse, are shown in Fig.~\ref{fig1}. 
We obtain $z=5.85(9)$ for the bimodal case and $z=6.00(10)$ for the Gaussian case. The plateau on the low velocity side indicates the fully adiabatic regime, 
while the straight line in these log-log plots show the universal scaling governed by the exponent $x$ in Eq.~(\ref{eq:power}). The value of $x$ extracted 
from the slope agrees very well with the expression in Eq.~(\ref{eq:power}). We have, thus, demonstrated dynamic scaling at the spin glass transition and 
its cross-over into the standard equilibrium finite-size scaling.

\begin{figure}
\includegraphics[width=7cm, clip=true]{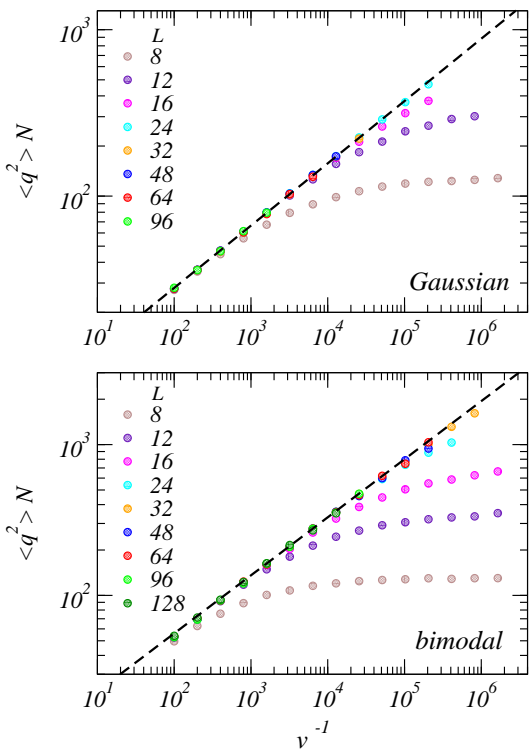}
\caption{(Color online) Data collapse onto the function $f_2$ in Eq.~(\ref{eq:dual_scaling}).
The linear regimes in these log-log plots are governed by the exponent $x$ in Eq.~(\ref{eq:power}) and the low-$v$
deviations from linearity are manifestations of the intrinsic length-scale $\xi_v$ approaching the system size $L$.}
\label{fig2}
\end{figure}

The second scaling function, $f_2$ in Eq.~(\ref{eq:dual_scaling}), suggests an easier way of extracting the exponent $x$: Graphing 
$\langle q^2 \rangle L^d$ versus $v^{-1}$ the scaling regime emerges, as we show in Fig.~\ref{fig2}. Note that Figs.~\ref{fig1} and \ref{fig2} 
show the same data sets (though Fig.~\ref{fig1} omits the high-$v$ points that do not collapse), which are only graphed differently 
according to either the low- or the high-velocity scaling expectation. In Fig.~\ref{fig2} the highest velocity is $v=0.01$, which
is not yet high enough to see the cross-over into the eventual $v$-independent behavior governed by the initial state.

The characteristic velocity $v_{_{KZ}}$ suggests a correlation length $\xi_v \sim v^{-1/(z+1/\nu)}$, and the $f_2$ scaling collapse 
should apply when $\xi_v \ll L$. Therefore, when observing $f_2$ scaling the system is effectively in the thermodynamic limit, and the deviation from a common 
scaling function in Fig.~\ref{fig2} signals $\xi_v$ becoming comparable to $L$. By line-fitting in the power-law regime for $L \ge 64$ we obtain $x=0.3851(45)$ 
for the bimodal case and $x=0.3745(66)$ for the Gaussian case. Using the known values of the static exponents we then obtain $z=5.82(7)$ for the bimodal 
case and $z=6.00(10)$ for the Gaussian case. 

The above analysis shows that the two scaling forms in Eq.~(\ref{eq:dual_scaling}), approaching the power-law regime either from low ($f_1$) or 
high ($f_2$) velocities, are mutually consistent and smoothly connected. One can obtain the power $x$ from either way of analyzing the data.
Working from the high-velocity end with $f_2$ may be more practical when studying a new case since high-velocity data can be generated faster 
and one can easily monitor the emergence of the scaling regime. It is, however, also comforting to observe the cross-over into the equilibrium form.

\textit{Correction to scaling}.---The above analysis did not need any scaling corrections, as the data for the larger system sizes follow the
expected forms very precisely despite the error bars being very small. It is nevertheless important to investigate possible effects of corrections to
scaling, which we do here using the high-velocity regime and the cross-over into the universal scaling regime. A natural extension of the
$f_2$-part of Eq.~(\ref{eq:dual_scaling}) is 
\begin{equation}
\langle q^2 \rangle L^d \sim a \hspace{1pt} v^{-x} (1 + b \hspace{1pt} v^{x'}),
\label{eq:f2_correction}
\end{equation}
where one would expect the subleading term with exponent $x'$ to accommodate the initial cross-over away from the pure power law into high-velocity 
scaling governed by the initial state. Using the data for $v \le 0.01$ until the finite-size effects set in, we obtain $x=0.376(17)$, 
$x'=0.34(11)$,  $a=10.4(7)$, $b=-0.4(2)$ from sizes $L \ge 32$ for the bimodal case and $x=0.373(25)$, $x'=0.74(34)$, $a=5(1)$, $b=-1(3)$ using sizes 
$L \ge 48$ for the Gaussian case. The small values and large uncertainties in $b$ show that the corrections are statistically marginal and their only 
effect in practice larger error bars on $z$ due to the larger number of fitting parameters---we obtain $z=6.0(3)$ for the bimodal case and 
$z=6.0(4)$ for the Gaussian case. Since no corrections are needed to describe an extensive region of power-law scaling and adiabatic cross-over, 
we regard our results obtained without corrections in Fig.~\ref{fig1} as our best estimates of $z$.

\textit{Discussion}.---The non-equilibrium MC simulations and accompanying scaling analysis we have used here in large-scale studies of 3D Ising spin glasses 
demonstrate the utility of this method for highly frustrated systems. We have used existing knowledge of the $T_c$ values and static exponents of the 
systems studies, and the remarkably good fits with the dynamic exponent $z$ as the only adjustable parameter in Fig.~\ref{fig1} reinforce the reliability of 
the other parameters used as input. The dynamic MC scheme can also be used to extract critical points and static exponents, as has recently been done 
for quantum models in Refs.~\cite{liu:13,liu:14b}. We note that the advantage demonstrated here of circumventing difficulties due to critical slowing 
down only applies, currently, to classical systems, where the dynamics of interest is the dynamics of the simulation method itself, while in quantum 
systems one is typically interested in Hamiltonian dynamics with no direct relation to the updating scheme used in the simulation. Our results for the 
Ising glasses provide strong support for a universal dynamic exponent with single-spin Metropolis MC updates, and one would naturally expect this 
to extend to any local dynamics.

\begin{acknowledgments}
We thank Arnab Das and Roderich Moessner for helpful discussions. The work of CWL, AP, and AWS is supported by the NSF under grant 
No. PHY-1211284. APY acknowledges support from a Gutzwiller Fellowship at the Max Planck Institute for the Physics of Complex Systems, the Humboldt Foundation, 
and the NSF through grant DMR-1207036.
\end{acknowledgments}

\end{document}